# GaMnAs: layers, wires and dots

Janusz Sadowski[1,2]

[1] *Institute of Physics, Polish Academy of Sciences, al. Lotników 32/46, 02-668 Warszawa, Poland*
[2] *MAX-Lab, Lund University, 221 00 Lund, Sweden*

Thin layers of GaMnAs ferromagnetic semiconductor grown by molecular beam epitaxy on GaAs(001) substrates were studied. To improve their magnetic properties the post-growth annealing procedures were applied, using the surface passivation layers of amorphous arsenic. This post growth treatment effectively increases the ferromagnetic-to-paramagnetic phase transition temperature in GaMnAs, and provides surface-rich MnAs layer which can be used for formation of low dimensional structures such as superlattices. If the surface rich MnAs layer consists of MnAs dots, then it is possible to grow Mn-doped GaAs nanowires.

## 1. Introduction

After the discovery of ferromagnetism in GaMnAs by H. Ohno et. al. in 1996 [1] this compound became the most comprehensively studied ferromagnetic semiconductor with the ferromagnetic phase induced by interactions between spins of Mn ions and free carriers (valence band holes). Mn in GaAs lattice constitutes a source of uncompensated spins and at the same time is an efficient p-type dopand, providing concentration of holes in the range of $10^{20}$ – $10^{21}$ $cm^{-3}$ for the Mn content of a few percent. The preparation of GaMnAs requires low temperature molecular beam epitaxial (MBE) growth. Use of very low growth temperatures, much lower than usually applied for GaAs, induces formation of compensating defects such as As antisites and Mn interstitials. Both defects are double donors, decreasing the concentration of holes provided by $Mn_{Ga}$ acceptors [2, 3]. The concentration of the first defects can be minimized by appropriate choice of growth parameters [4], the latter can be removed by suitable annealing procedures applied after the MBE growth [5, 6]. By use of optimized growth and post-growth annealing procedures a ferromagnetic phase transition temperature ($T_c$) close to 180 K has been attained recently [7]. Further increase of $T_c$ will probably demand increasing Mn content above 10%, accompanied by sufficiently high concentration of holes. Recently it was reported that uniform

GaMnAs can be grown with Mn content of up to 20% [8-10], so progress in this direction cannot be excluded.

Besides GaMnAs in form of layers, also low dimensional structures such as quantum wires have been studied. So far they were fabricated from layered structures with use of e-beam lithography [11-13]. In addition to structures obtained this way, it is possible to get self-assembled one-dimensional structures (nanowires). They are generated due to the catalyzing properties of MnAs nano-islands formed at the GaMnAs surface during MBE growth above the phase separation threshold [14]. In GaAs:Mn nanowires with sufficiently small diameters GaAs occurs in the hexagonal (wurzite) phase [15]. So far GaMnAs in this structure has not been studied.

## 2. Basic properties of GaMnAs

### 2.1 Epitaxial growth of GaMnAs

GaMnAs can be grown by molecular beam epitaxy (MBE) on GaAs substrates when the substrate temperature during the growth process is much lower than usually applied for GaAs. In the latter case the MBE growth is performed with the substrate temperature typically in the range 590 – 640 $^{o}$C. For GaMnAs much lower temperatures must be used in order to avoid segregation of MnAs clusters (MnAs is a ferromagnetic metal with ferromagnetic to paramagnetic phase transition temperature of 40 $^{o}$C). The segregation of MnAs clusters is easily detectable in reflection high energy electron diffraction (RHEED). Fig. 1 shows an example of RHEED pictures from smooth, uniform GaMnAs film, and from a GaMnAs with surface segregated MnAs clusters.

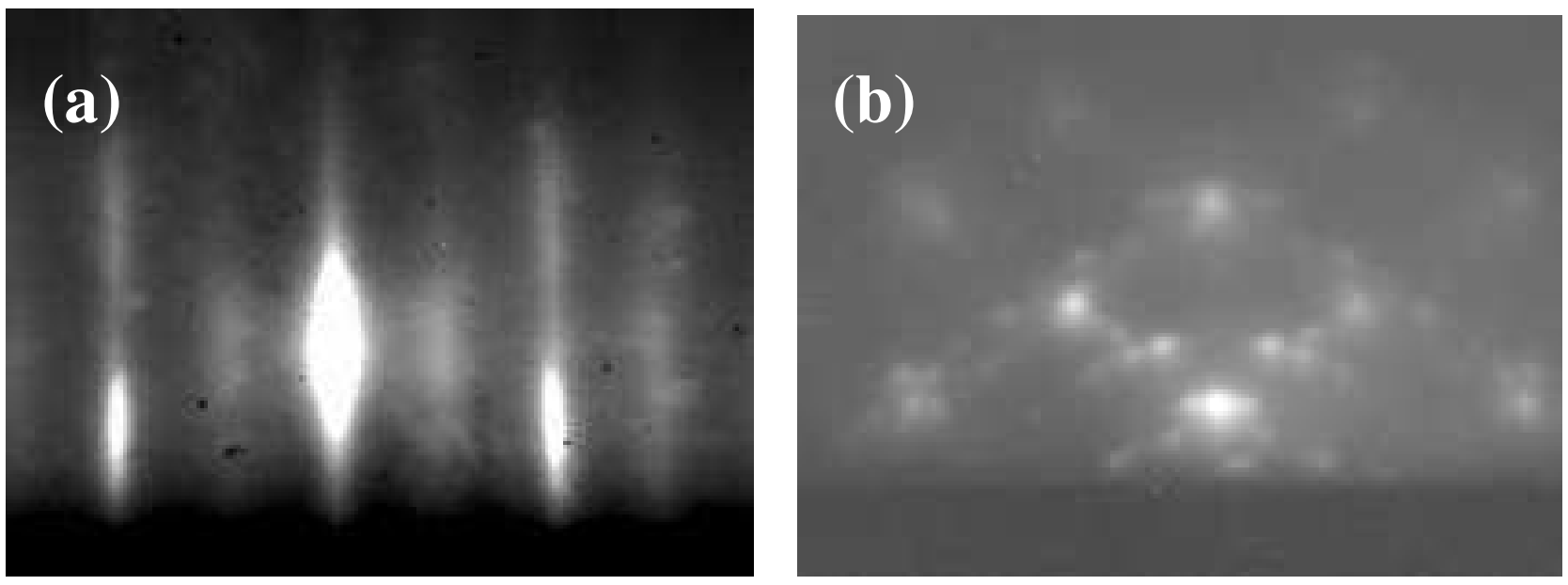

**Fig.1.** *RHEED diffraction images from the surface of smooth $Ga_{0.94}Mn_{0.06}As(001)$ layer (a) and $Ga_{0.94}Mn_{0.06}As$ with surface segregated MnAs islands. The direction of e-beam is parallel to the [-110] azimuth.*

In order to avoid segregation of MnAs the growth of GaMnAs with Mn content higher than 1 at. % must be performed below 300 $^{o}C$. The low growth temperature implies generation of structural defects such as As antisites ($As_{Ga}$) with maximum concentrations reaching 0.5 at.% [16]. $As_{Ga}$ is a deep donor in GaAs, so it compensates part of the $Mn_{Ga}$ acceptors. The presence of $As_{Ga}$ cannot be avoided under LT growth conditions. The concentration of these defects can be minimized by using suitable MBE growth conditions, i.e. maximum substrate temperature, slightly below the MnAs segregation threshold [17], and close to stoichiometric supply of the two elements, i.e. relatively low As/Ga flux ratio. A schematic diagram describing GaMnAs growth is shown in Fig.2.

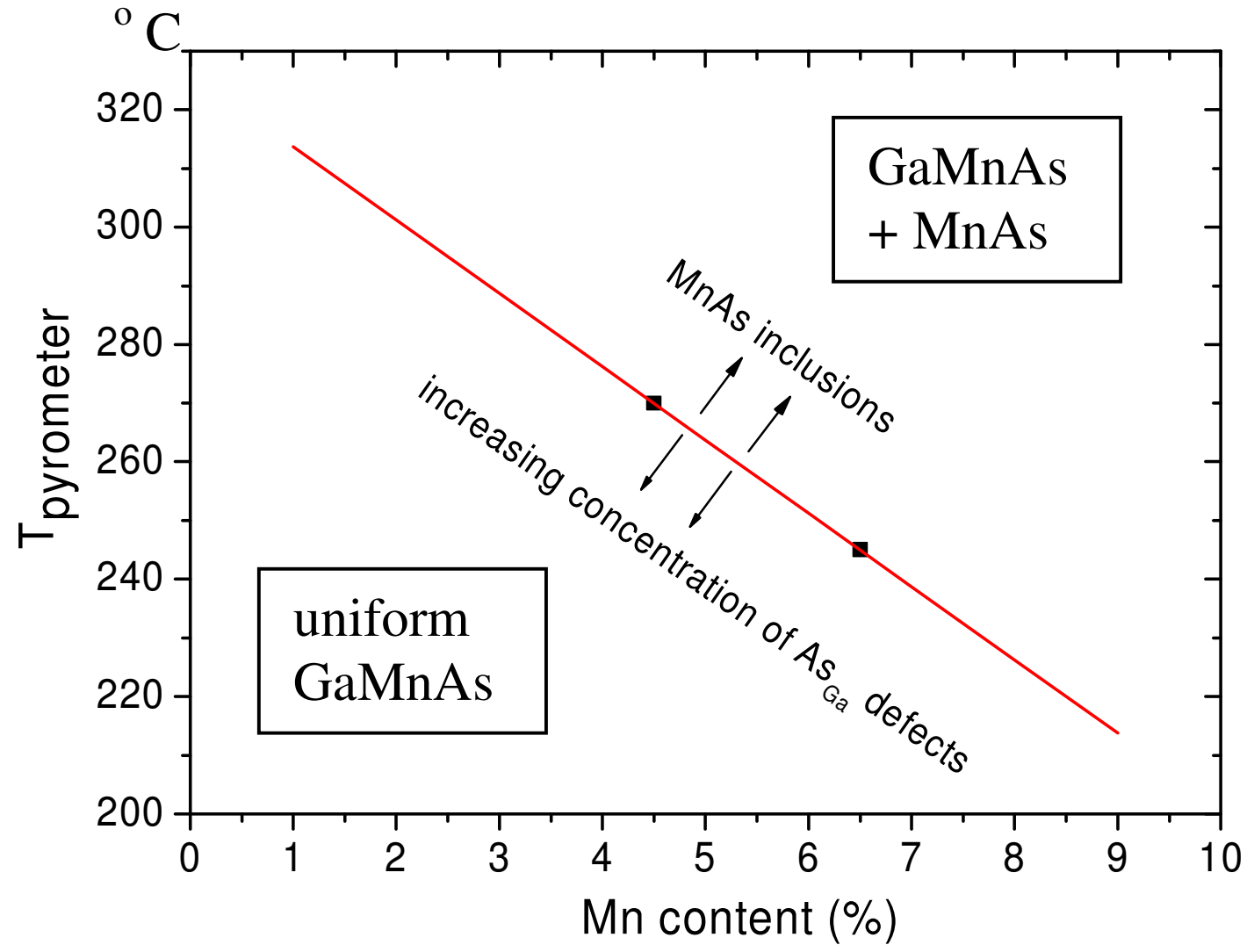

**Fig.2.** *Schematic diagram for GaMnAs growth. The optimum growth conditions are close to the line separating regions of single phase and MnAs segregation.*

Another defect, specific for GaMnAs, is Mn located at interstitial sites ($Mn_I$). It has been reported that up to 20% of a total Mn content can be located in interstitial sites [18]. Since $Mn_I$ also acts as a double donor, it considerably lowers the concentration of holes in GaMnAs, decreasing the temperature of paramagnetic to ferromagnetic phase transition.

### 2.2. Post-growth annealing

In contrast to $As_{Ga}$ the $Mn_I$ defects can be removed from the GaMnAs bulk by applying post-growth annealing. Due to the annealing the weakly bounded Mn interstitials diffuse to the GaMnAs surface and are passivated either by oxidation, or by binding with another reactive element [19], such as As deposited on GaMnAs surface as thick amorphous layer directly after the MBE growth. The first method requires very long annealing times (up to 100 hrs) and gives considerable increase of $T_c$ (from 60 – 100 K for as grown GaMnAs up to 170 – 180 K after annealing). The post-growth annealing method using amorphous As passivation layer allows to use much shorter annealing times, usually 1 – 3 hrs, and for sufficiently thin layers the surface of the annealed GaMnAs is well ordered and suitable for further epitaxial growth. The idea of this annealing method is schematically presented in Fig. 3.

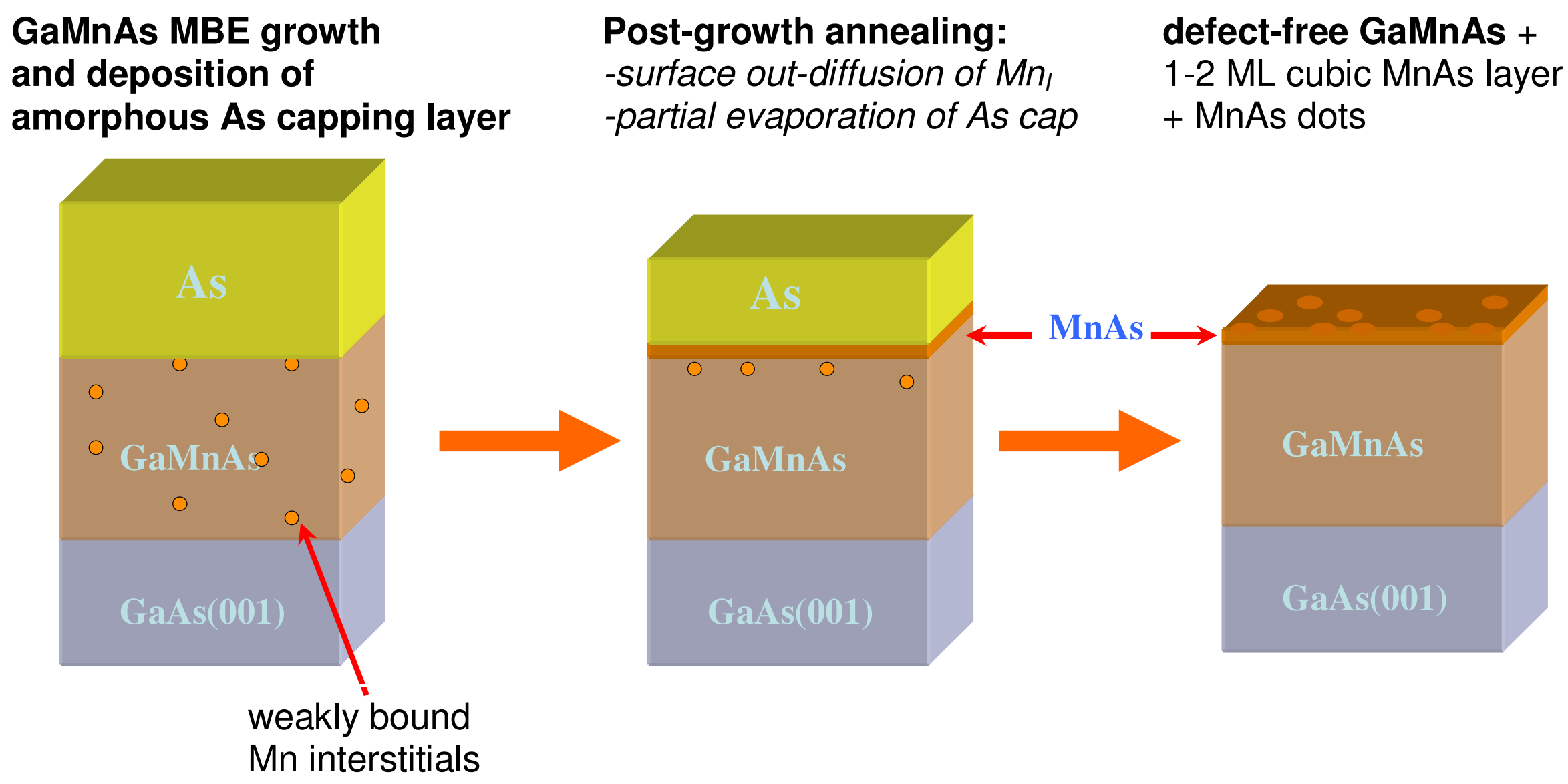

**Fig.3.** *Procedure of removing Mn interstitials from GaMnAs by post-growth annealing with passivating amorphous arsenic layer.*

The high efficiency of the annealing process is confirmed by magnetization measurements (see Fig. 4), which show a remarkable increase of $T_c$ from about 30 to about 130 K after 3hrs of annealing.

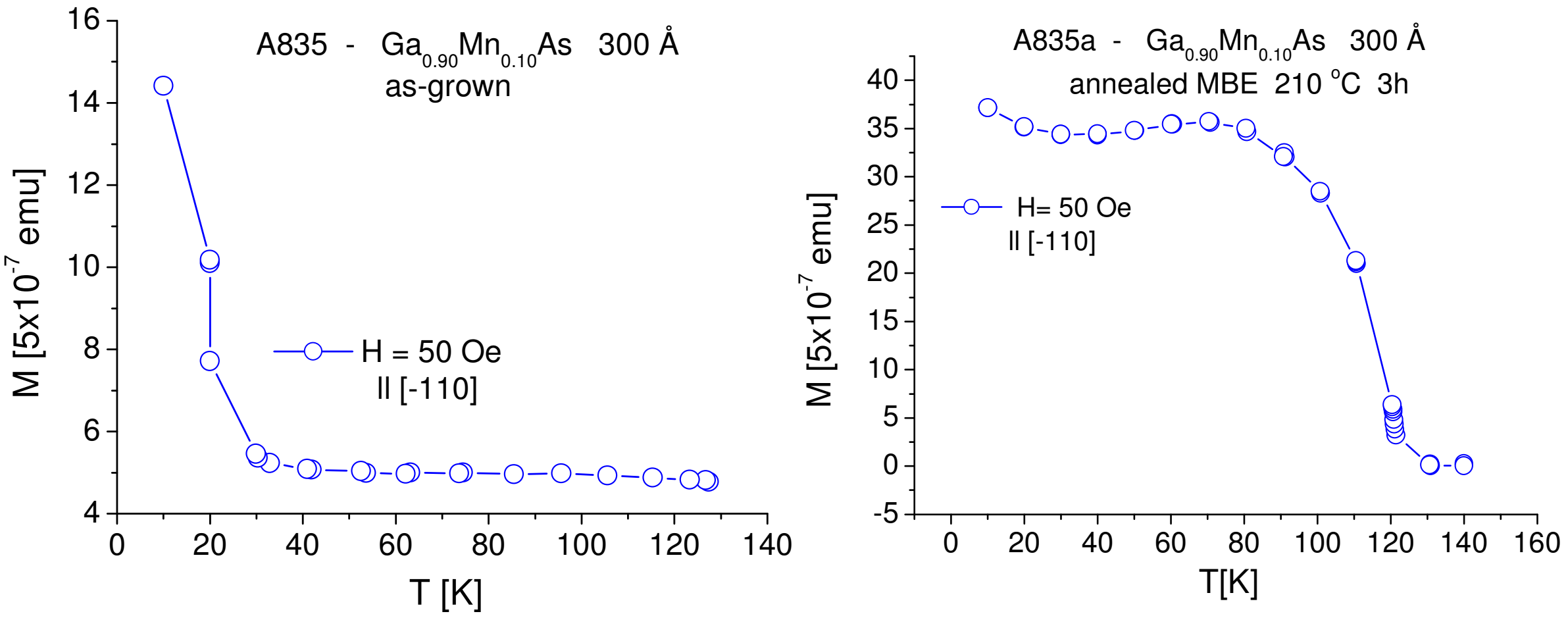

**Fig.4.** *Temperature dependence of magnetization measured by SQUID for a 300 Å thick $Ga_{0.9}Mn_{0.1}As$ layer before annealing (left panel) and after annealing with passivating As layer (right panel).*

Since Mn interstitials are double donors partially compensating $Mn_{Ga}$ acceptors, the annealing increases the concentration of holes 2-3 times [20]. The saturation moment also increases significantly, since Mn interstitials tend to couple antiferromagnetically with Mn at Ga sites, making part of $Mn_{Ga}$ atoms magnetically passive.

Besides magnetic and transport properties, post-growth annealing also affects the structural parameters of GaMnAs: Mn causes a slight increase of the GaAs lattice constant. According to the theoretical model by Masek et al. [21] the main contribution to the lattice constant increase is due to Mn interstitials. Thus it is expected that upon annealing the GaMnAs lattice constant will be reduced. Fig. 5 shows results of X-ray diffraction measurements for a 700 Å thick $Ga_{0.94}Mn_{0.06}As$ layer. The lattice constant reduction can be seen as shifts of diffraction peaks from GaMnAs layer to the higher diffraction angles. Already after 1 h annealing there is a remarkable lattice constant reduction. The effects saturates after annealing for 3 h. Annealing for longer time (30 hrs) doesn't lead to further lattice constant reduction. The broadening of the GaMnAs diffraction peaks is due to the small layer thickness.

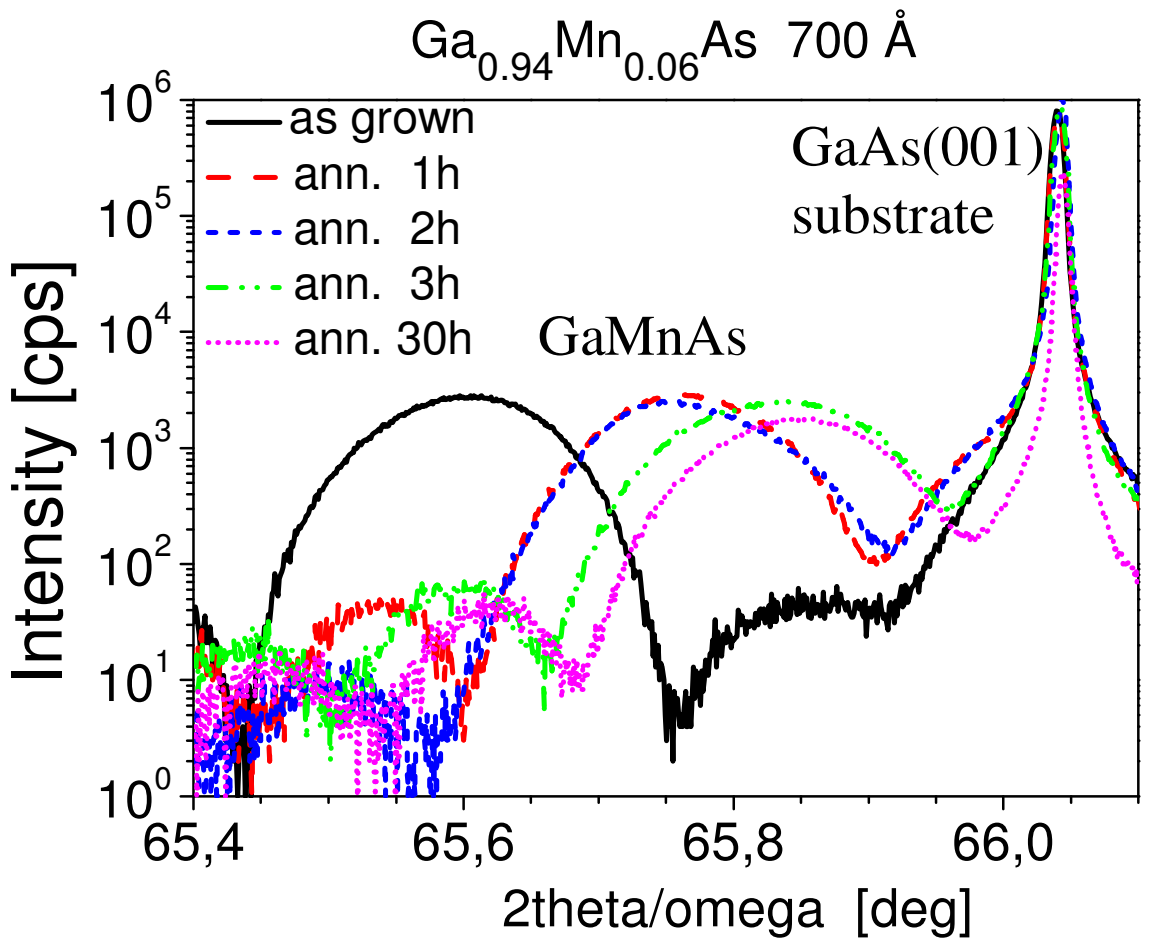

**Fig.5** *(004) Bragg reflection from 700 Å thick $Ga_{0.94}Mn_{0.06}As$ layer grown on GaAs(001) substrate before annealing, and after post growth annealing under As capping at 190 $^oC$ for 1, 2, 3 and 30 hrs.*

It was observed that post-growth annealing of GaMnAs is only efficient for the layers which are not too thick. The thickness limit for the annealing process is close to 1000 Å [6]. It has been tentatively explained by formation of a surface layer blocking further out-diffusion of remaining Mn interstitial atoms. In case of annealing in air this is not well understood since the structure of GaMnAs surface oxidized layer is not well known. In case of annealing in high vacuum, with the presence of As, annealing leads to formation of a thin (1-2 ML) MnAs layer [20] in the zinc-blende phase. For higher amount of Mn interstitials formation of MnAs dots was detected [22]. Fig. 6 shows comparison of surface morphologies of 200 Å thick and 500 Å thick GaMnAs films after annealing with As capping layer. At the end of annealing process the As capping was completely desorbed, revealing the GaMnAs surface morphology.

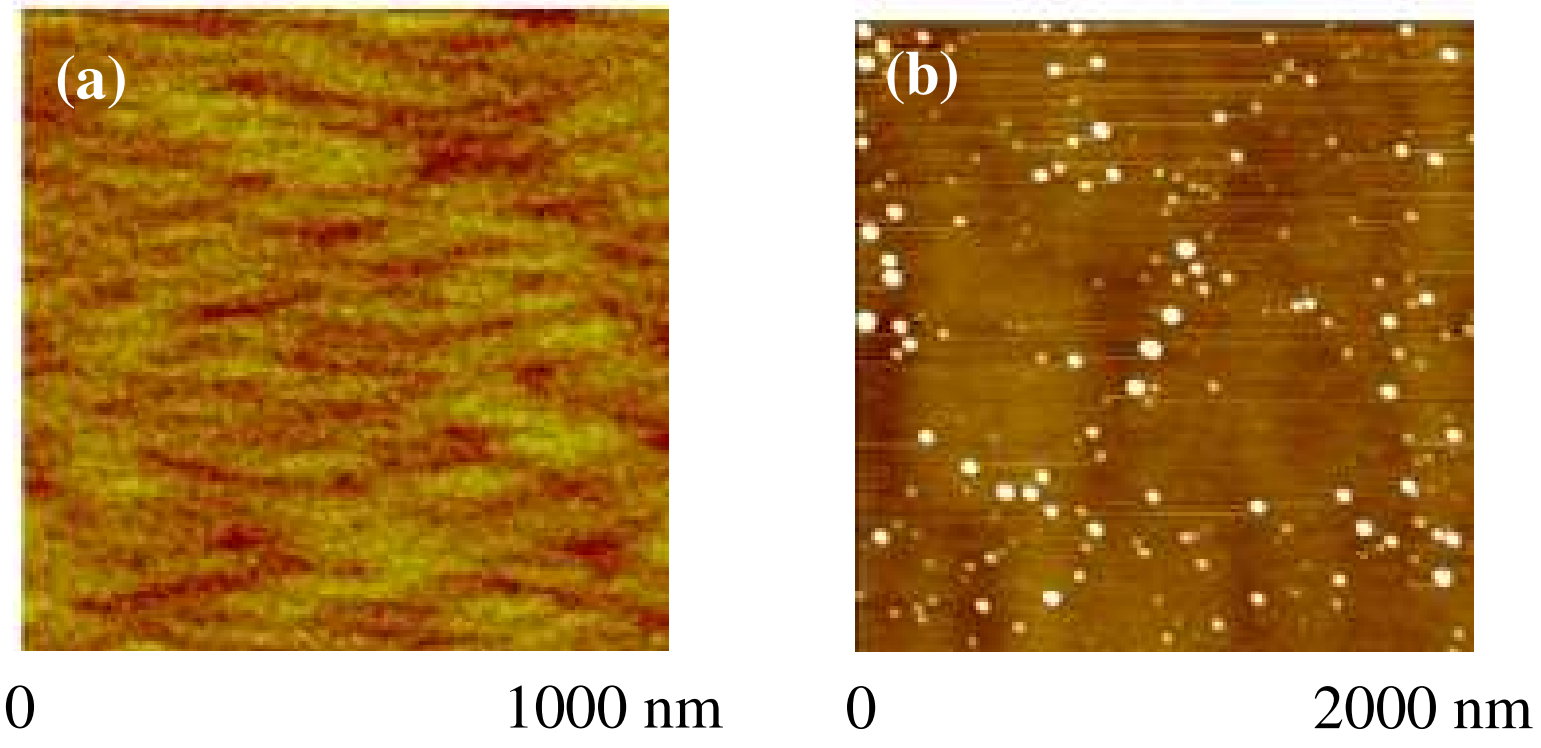

**Fig. 6.** *AFM images of (a) 200 Å, (b) 500 Å thick* $Ga_{0.94}Mn_{0.06}As$ *after annealing with As capping. At the end of annealing the As capping layer was completely desorbed.*

The sequence of GaMnAs growth, As capping and annealing in the MBE system can be repeated leading to the GaMnAs/MnAs(zinc-blende) superlattice structure. Fig 7 shows the results of XRD and magnetization measurements of superlattice structure with [$Ga_{0.93}Mn_{0.07}As$(75 Å) /MnAs] sequence repeated 8 times. Satellite peaks in X-ray diffraction clearly prove the presence of a superstructure.

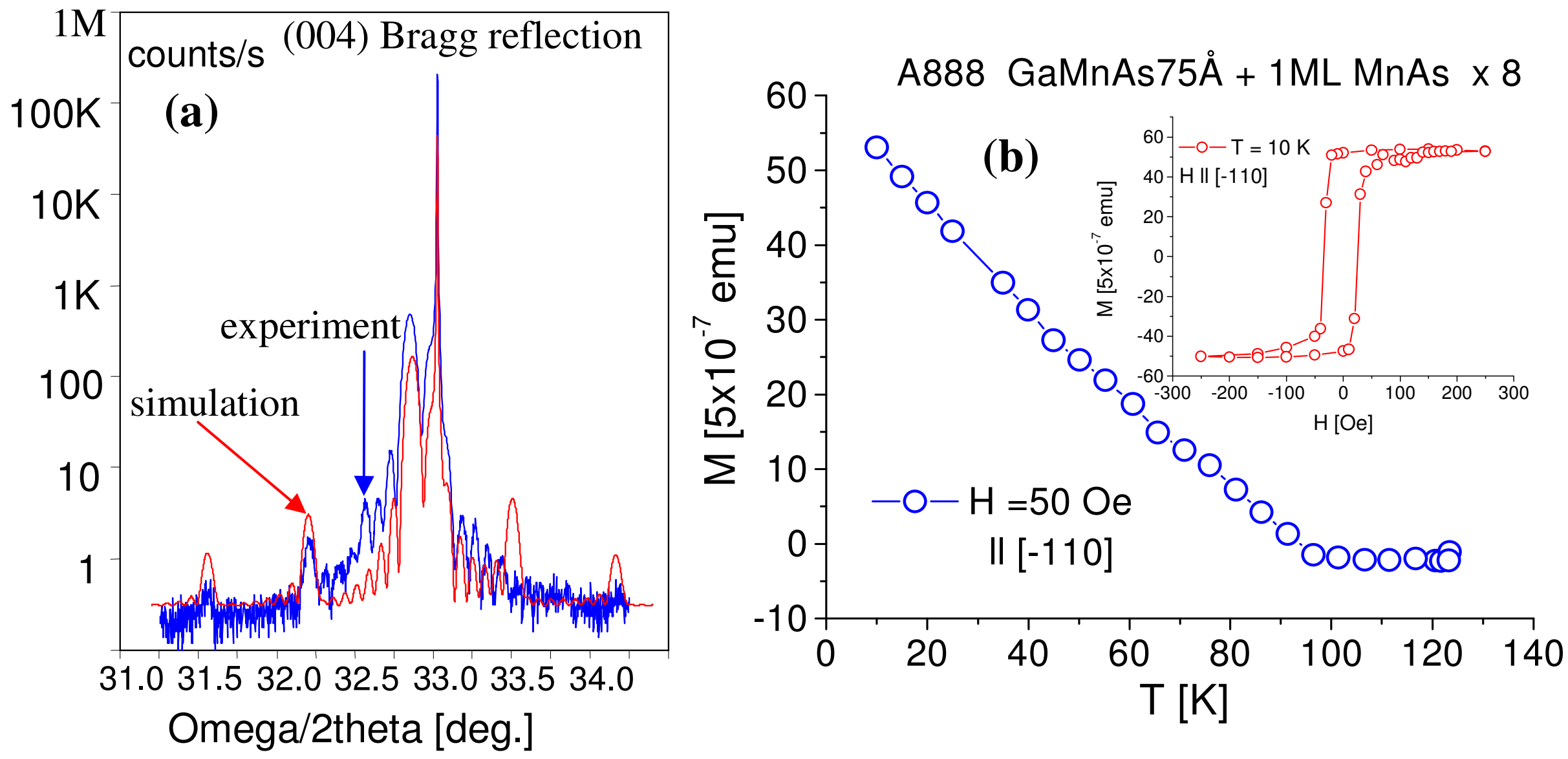

**Fig. 7.** *(a) X-ray diffraction picture from [$Ga_{0.93}Mn_{0.07}As$(75 Å)/MnAs(1 ML)] x 8 superlattice fabricated by subsequent GaMnAs MBE growth, As capping and annealing; (b) temperature dependence of magnetization and hysteresis loop (inset).*

The shape of the M(T) curve shown in Fig. 7.b. is rather untypical for GaMnAs, this may be due to the interactions between two magnetic systems i.e. GaMnAs and MnAs layers.

As in the case of annealing in air, annealing with As passivation layer also shows a reduced efficiency for layers thicker than about 1000 Å. This can be understood as a result of blocking further $Mn_I$ out-diffusion by the surface MnAs layer (or Mn-rich oxidized surface layer in case of annealing in air [6]). To get around this limitation it is obvious that the reacted layer has to be removed. This procedure was recently demonstrated by Olejnik et. al. [7] for samples annealed in air. By chemical etching of the reacted surface and repeated annealing a record Tc of 180 K was achieved.

### 3. GaMnAs layers under tensile strain

GaMnAs grown on GaAs(001) substrate is in a compressive strain state, which implies that the easy magnetic axis is in-plane (can also be out of plane in some narrow range of GaMnAs parameters such as Mn content and concentration of holes). In many cases it is desirable to use layers with perpendicular easy magnetization axis. This can be achieved in GaMnAs under tensile strain, i.e. for growth on a surface with larger lattice constant. Using GaAs as initial substrate, a buffer with a larger lattice constant is obtained by deposition of a thick, relaxed InGaAs layer. Figure 8 shows the results of XRD measurements for 500 Å thick $Ga_{0.925}Mn_{0.075}As$ layer deposited on a 1500 nm thick $In_{0.07}Ga_{0.93}As$ buffer. Detailed information concerning strain state i.e. in plane and out of plane lattice parameters of both InGaAs and GaMnAs layers was obtained from reciprocal space maps of asymmetrical (224) Bragg reflections. Fig. 8b shows a reciprocal space map for the same sample as shown in Fig. 8a. A 100% relaxation line (solid line) shows that InGaAs buffer is only partially relaxed. The GaMnAs is, however, completely strained to the InGaAs buffer (i.e. it has the same in plane lattice constant as InGaAs) as shown by the dashed line.

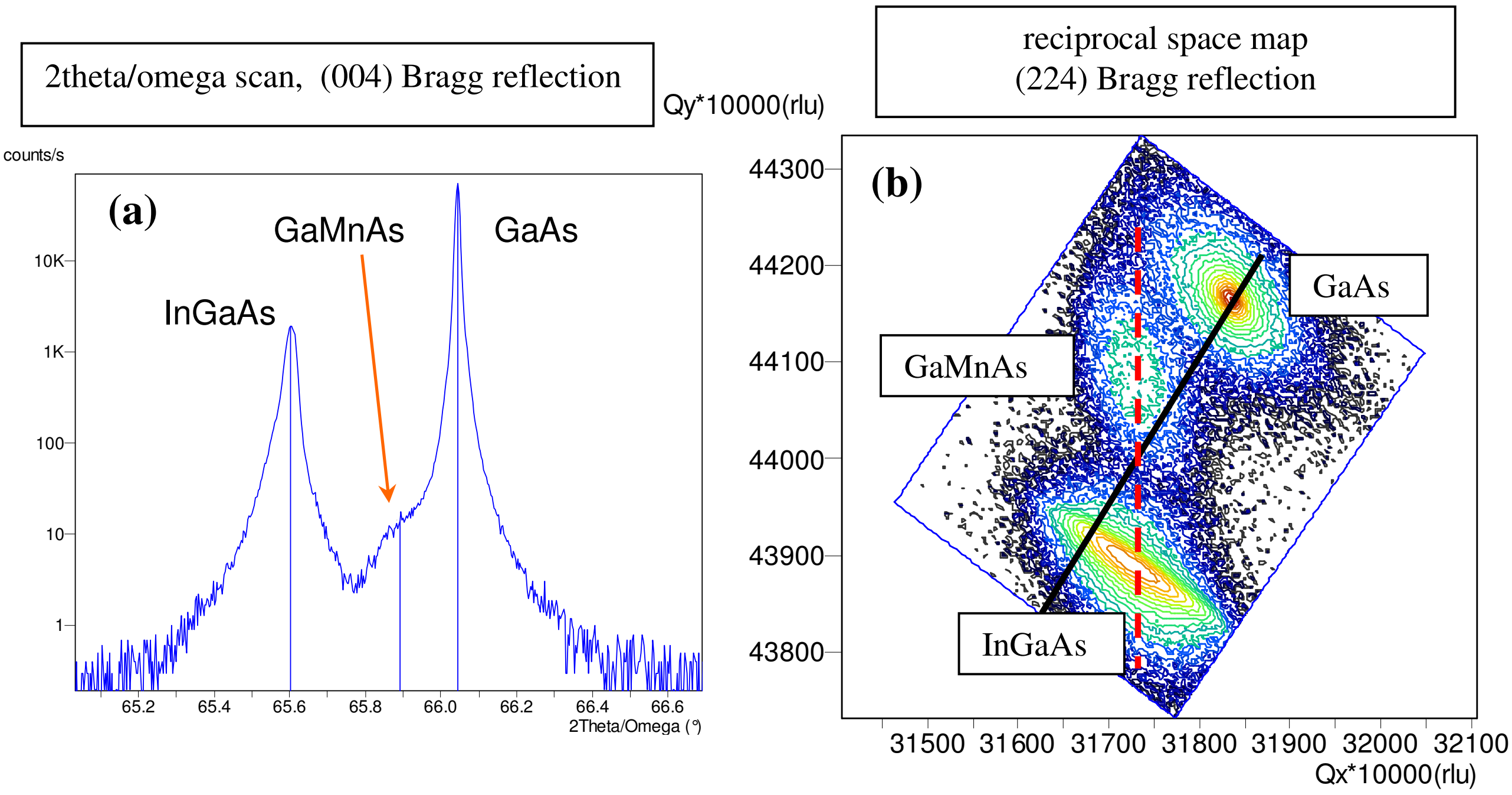

**Fig.8.** *(a) 2theta/omega scan around (004) Bragg reflection from 500 Å thick $Ga_{0.925}Mn_{0.075}As$ layer deposited on 1500 nm thick $In_{0.07}Ga_{0.93}As$ buffer; (b) reciprocal space map of asymmetrical (224) Bragg reflection*

The sample used for XRD measurements shown in Fig. 8 was subjected to the As capping and post growth annealing in the MBE system for 1h 45 min at 210 $^{o}$C. The magnetic properties of this sample i.e. temperature dependence of magnetization and hysteresis loop are shown in Fig. 9a. For comparison, Fig 9b shows results of the same measurements of the twin sample, grown in the same MBE growth run, but without the InGaAs buffer. The higher Curie temperature of sample grown on InGaAs buffer proves the higher efficiency of the annealing process in this case.

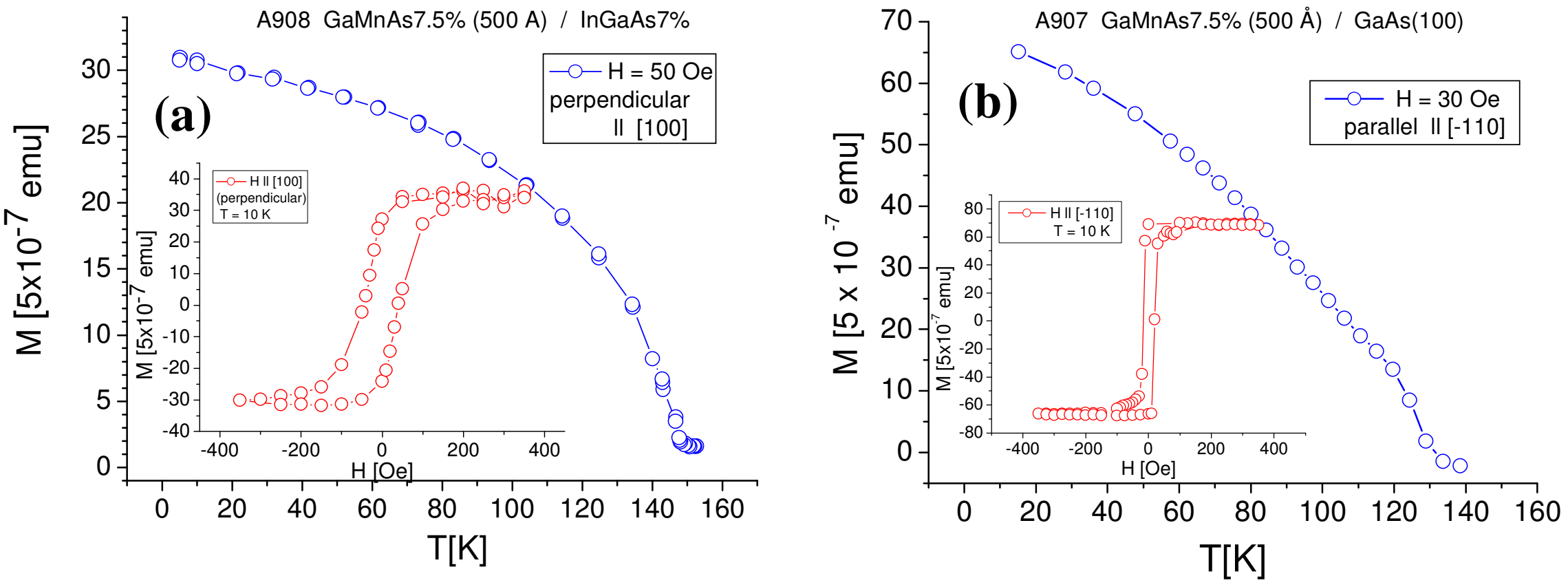

**Fig.9.** *(a) M(T) curve and hysteresis loop for $Ga_{0.925}Mn_{0.075}As$ layer deposited on 1500 nm thick $In_{0.07}Ga_{0.93}As$ buffer. The sample has a perpendicular easy magnetization axis; (b) M(T) curve and hysteresis loop for identical GaMnAs layer grown without InGaAs buffer, having an in plane easy magnetization axis. The samples were annealed simultaneously with an As passivation layer directly after the MBE growth.*

The higher annealing efficiency of GaMnAs grown on InGaAs buffer may be due to the surface structure, which in case of InGaAs buffer layers exhibits a micrometer scale corrugation in the form of cross-hatched wedges parallel to [-110] and [110] crystallographic directions.
Fig. 10 shows an AFM image of the 500 Å thick $Ga_{0.94}Mn_{0.06}As$ layer deposited on $In_{0.11}Ga_{0.89}As$ buffer. The surface structure of thick InGaAs buffer is also present at the surface of GaMnAs layer.

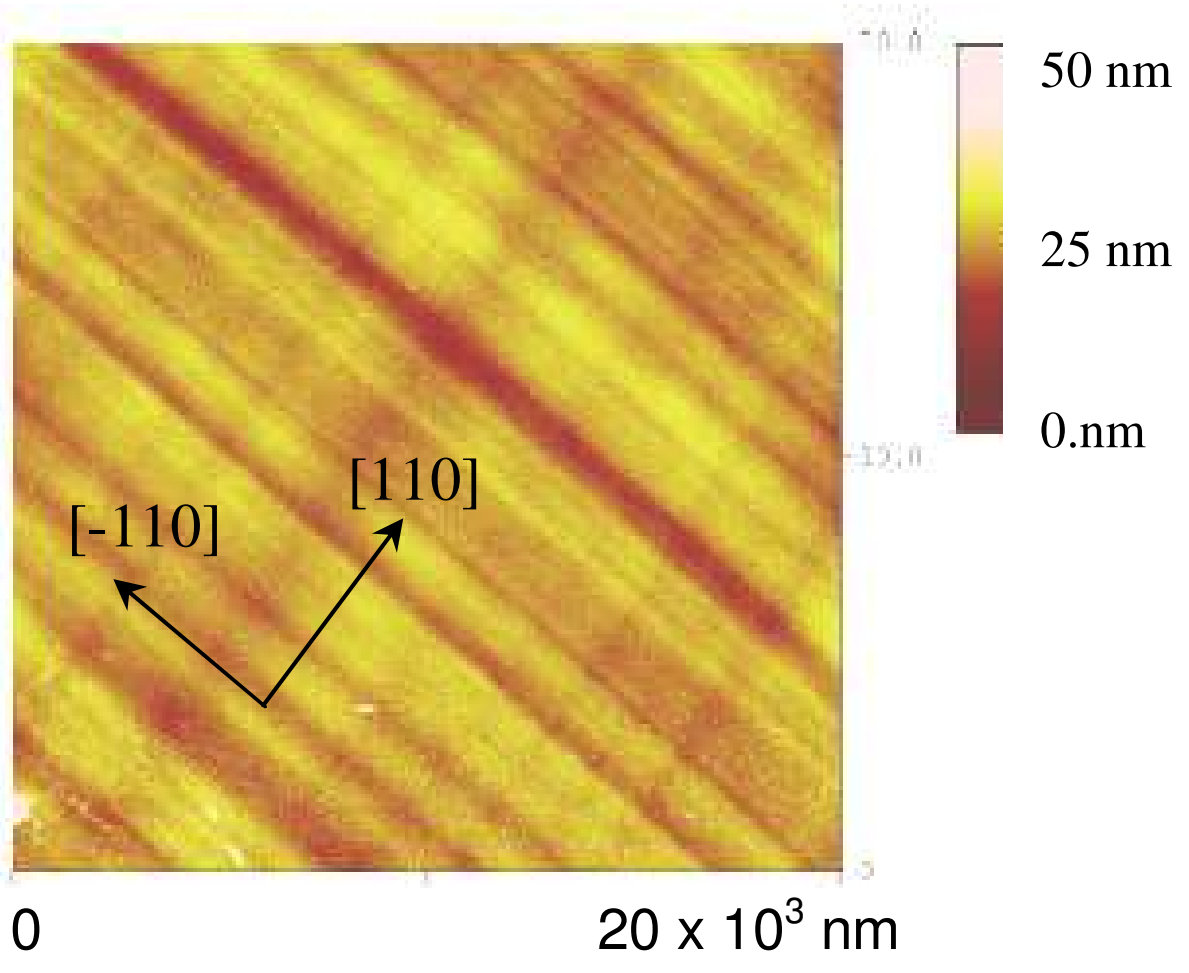

**Fig. 10.** *AFM image of 500 Å thick $Ga_{0.94}Mn_{0.06}As$ layer deposited on $In_{0.11}Ga_{0.89}As$ buffer. Cross-hatched surface wedges parallel to [110] and [-110] crystallographic directions are about 50 Å deep.*

## 4. Surface MnAs clusters as nanowire growth catalyst.

If the MBE growth is performed at too high substrate temperature or too high Mn flux MnAs clusters are formed at the surface of GaMnAs layer (see Fig.11). As observed by the author [14], continuing GaMnAs MBE growth in the presence of surface MnAs clusters, i.e. at phase segregation conditions, leads to formation of nanowires.

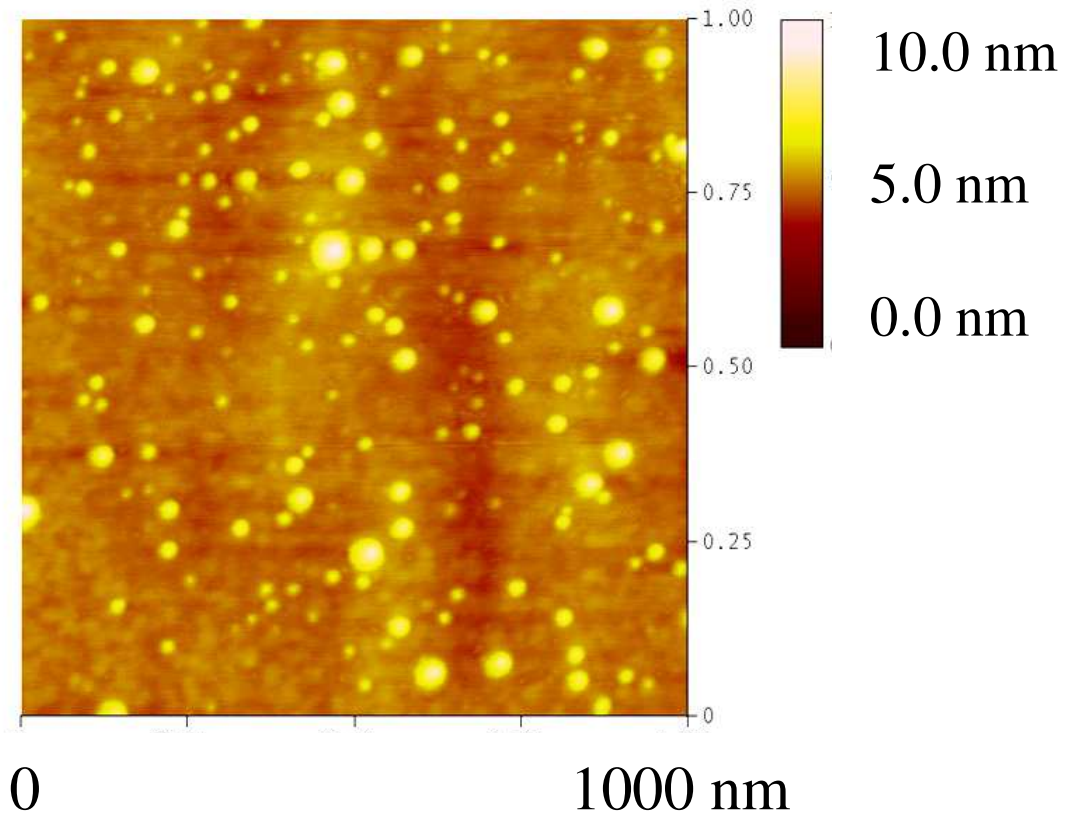

**Fig.11.** *AFM image of GaMnAs surface with segregated MnAs dots.*

In such a case the growth mode changes from 2-dimensional layer-by-layer to 3D growth. The evolution of RHEED diffraction images during growth of nanowires is shown in Fig. 12

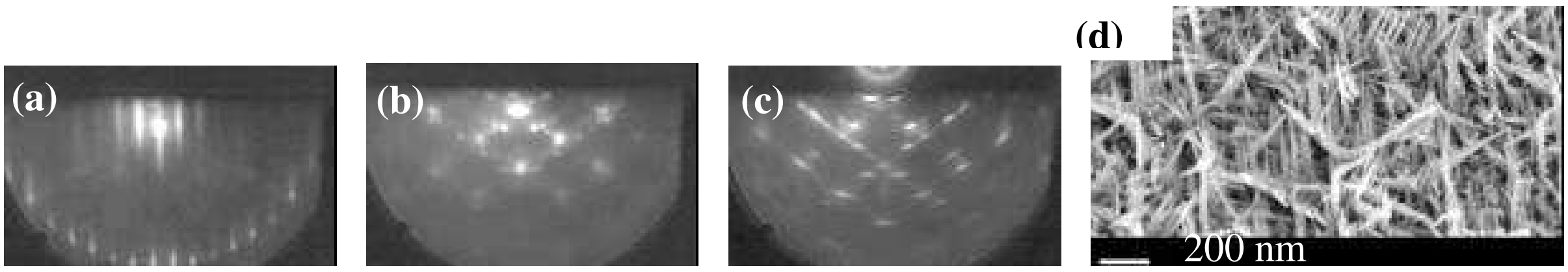

**Fig.12**. *(a) – (c) RHEED diffraction images recorded during 3-D growth of GaMnAs nanowires. (a) initial GaAs(001) surface in [110] azimuth, (b) and (c) diffraction images after 6 and 60 min of growth, respectively. (d) SEM picture from the final sample grown for 60 minutes*

The Mn flux supplied during the MBE growth presented in Fig. 12 corresponds to a Mn content of 6 at.% in uniform GaMnAs. For this large Mn flux the density of nanowires is very high, probably due to the catalyzing action of MnAs clusters generated during the whole growth process both at the surface between nanowires and on their sides. This latter process leads to formation of branches. Lowering the Mn flux during the NW growth reduces the density of NWs, as shown in the SEM pictures in Fig. 13.

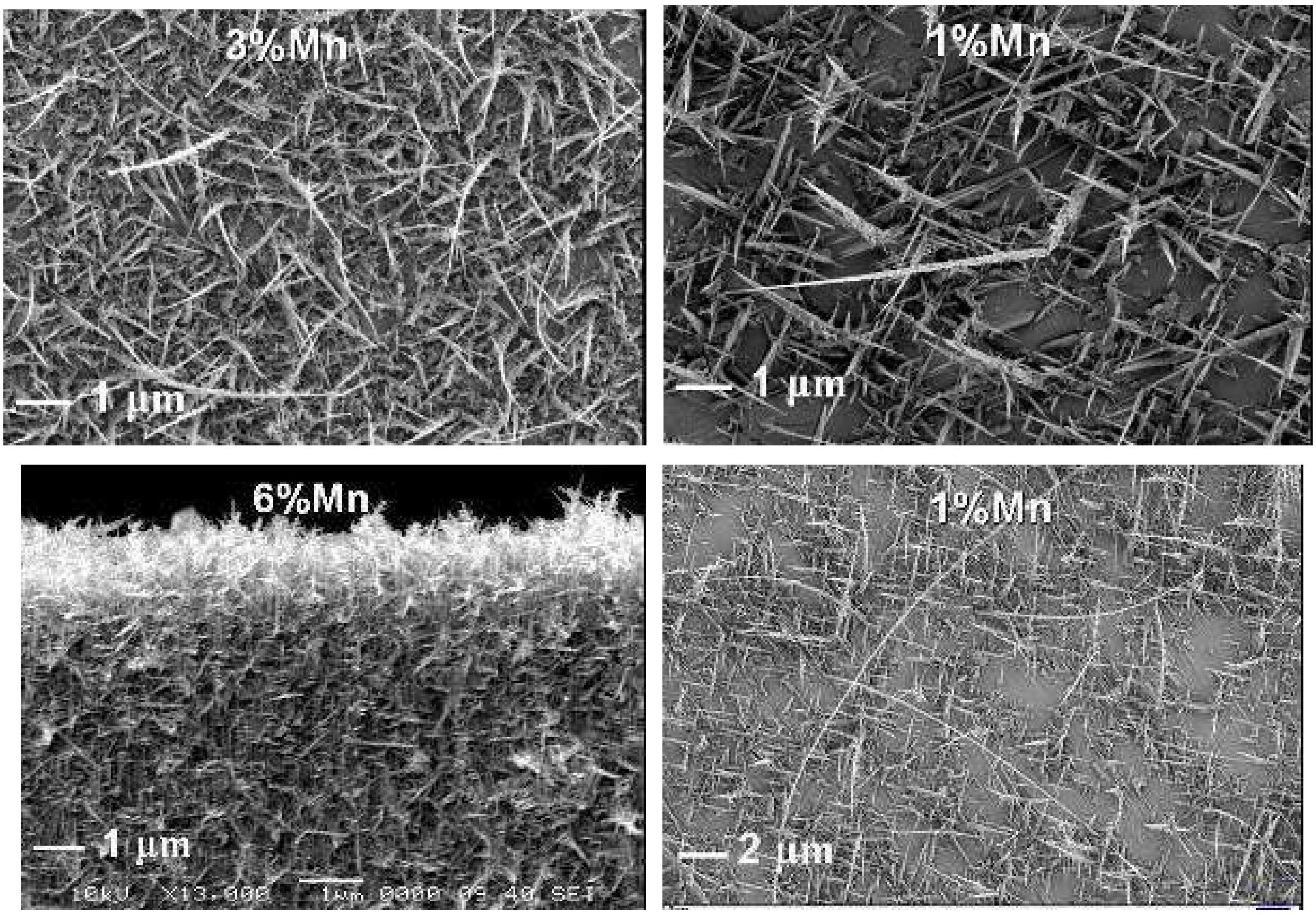

**Fig. 13.** *SEM pictures of GaMnAs nanowires grown on GaAs(001) substrate at different Mn fluxes.*

Fig. 14 shows SEM pictures of individual nanowires removed from the sample grown with 1% Mn and placed on a Si wafer.

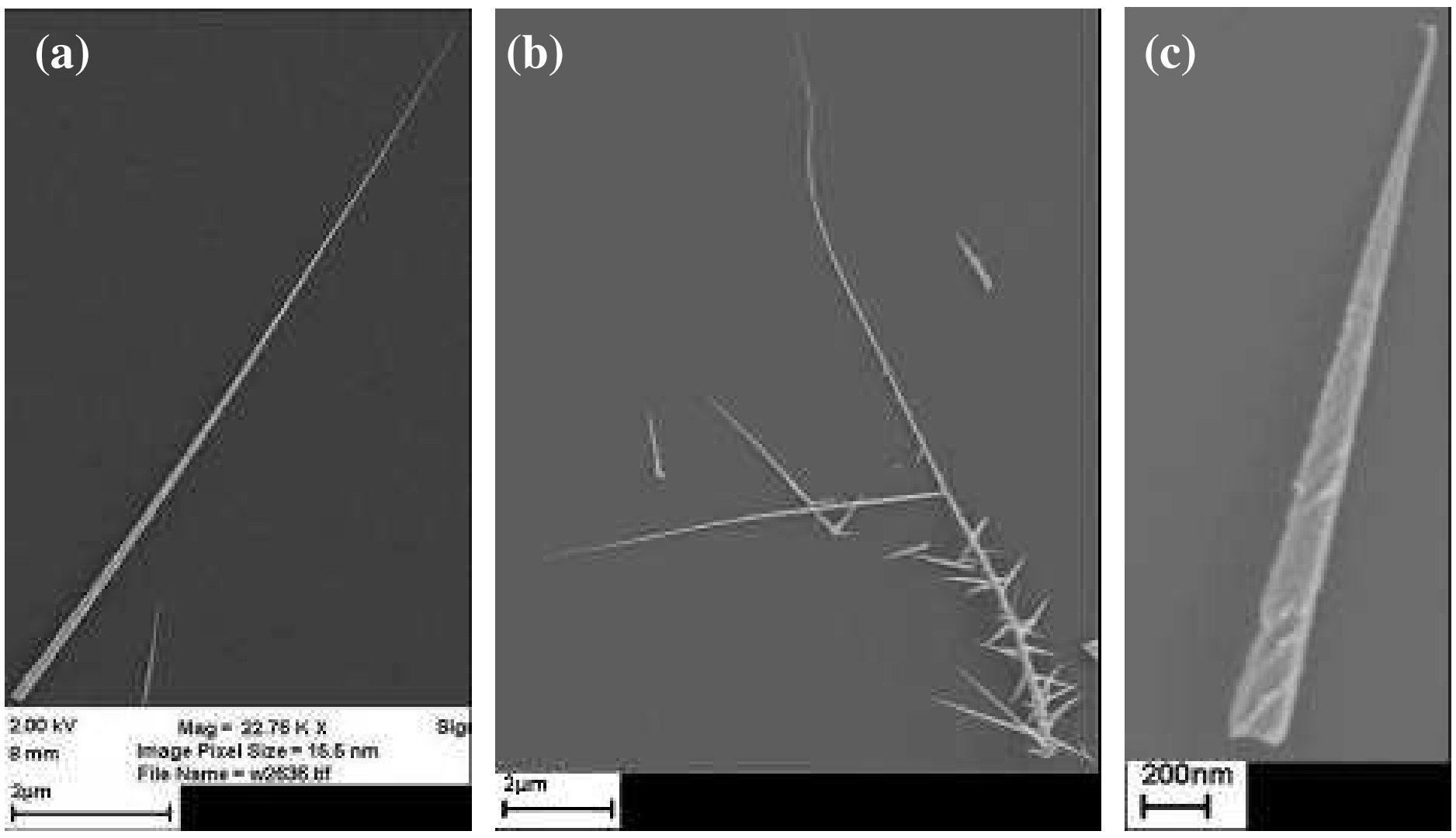

**Fig. 14.** *Three types of nanowires separated from the sample grown with 1% Mn (a) long nanowire without branches, (b) long nanowire with branches, (c) short, tapered nanowire*

There are still open questions concerning the properties of nanowires obtained in the GaMnAs MBE growth process at the MnAs segregation conditions. The as-grown samples with high density of nanowires were found to exhibit ferromagnetic properties [14], but the magnetism of individual nanowires still remains to be investigated.

**Conclusions.**

Post-growth annealing of GaMnAs with the presence of an As passivating layer has been used for removing Mn interstitials. Annealing can be done in the MBE system directly after the MBE growth. It was shown that this post-growth treatment effectively increases ferromagnetic to paramagnetic phase transition temperature in GaMnAs. The efficiency of the procedure is higher for GaMnAs layers grown on InGaAs buffers. As a side effect of the annealing a MnAs-rich surface is formed, consisting of either a very thin continuous layer of zinc-blende MnAs, or MnAs dots. In the first case it is possible to grow heterostructures containing GaMnAs and thin cubic MnAs. The latter case can be used to develop GaMnAs nanowires, where MnAs islands catalyze the nanowire growth.

## Acknowledgements

The author would like to thank Dr. J. Z Domagała from the Inst. of Physics Polish Academy of Sciences, Warsaw for the extensive collaboration in X-ray diffraction characterization of GaMnAs, E Łusakowska for AFM measurements, Dr. H. Shtrikmann from Weizmann Institute, Rehovot for SEM measurements of nanowires, and prof. J. Kanski from CTH, Göteborg for valuable discussions and careful reading of the manuscript. The GaMnAs project at MAX-Lab, Lund is partially financed by the Swedish Research Council (VR). The financial support from EADS (France) research project coordinated by prof. S. Charar and Dr. F. Terki, from GES Université Montpellier II; and access to the Braun Center for Submicron Research at the Weizmann Institute, Rehovot financed through the EC project RITA-CT-2003-506095 WISSMC is greatly acknowledged.